# A low-temperature setup for lock-in technique based dynamic magnetoelectric coupling measurements

Balwant Singh Chauhan[1], Priyanka Sharma[2], Rie Y. Umetsu[3,4], Ratnamala Chatterjee[1*]

*[1] Department of Physics, Indian Institute of Technology, Delhi, New Delhi 110016, India*

*[2]Department of Physics, Government Girls Polytechnic, Bulandshahr, Uttar Pradesh 203131, India*

*[3]Institute for Materials Research, Tohoku University, 2-1-1 Katahira, Aoba-ku, Sendai 980-8577, Japan*

*[4]Center for Science and Innovation in Spintronics, Tohoku University, 2-1-1 Katahira, Aoba-ku, Sendai 980-8577, Japan*

*Email: ratnamalac@gmail.com.*

**Abstract:**

Magnetoelectric (ME) phenomena in emerging material classes, such as two-dimensional van der Waals (vdW) magnets and Single-Molecule Magnets (SMMs), hold immense promise for next-generation cryogenic memory and quantum technologies. However, ME coupling in these systems predominantly manifests at low temperatures, making sensitive, cryo-compatible ME characterization techniques critical. To address this requirement, we report the design, validation, and performance of a custom closed-cycle refrigerator-based setup for dynamic lockin ME coupling measurements across 20-300 K under *dc* magnetic fields up to 7.5 kOe. Key design considerations for mitigating parasitic inductive background signals are also presented. The setup was validated on a $CoFe_2O_4$-$BaTiO_3$ (CFO-BTO) particulate composite, reproducing the characteristic room-temperature butterfly ME loop with a maximum ME coefficient of 0.23 mV/cm·Oe at ~3 kOe. Temperature-dependent measurements resolved ME anomalies at 200 K and 280 K, coinciding with the rhombohedral-orthorhombic and orthorhombic-tetragonal structural transitions of $BaTiO_3$, and were corroborated by simultaneous dielectric measurements on the same sample without cryostat reconfiguration. The instrument enables reliable ME and dielectric characterization down to 20 K, making it well suited for probing weak magnetoelectric coupling and phase transitions in multiferroic composites and quantum materials.

## 1. Introduction

The magnetoelectric (ME) effect, in which an applied magnetic field generates electric polarization (direct ME effect), or an applied electric field modulates magnetization (converse ME effect), has attracted considerable interest owing to its potential applications in multifunctional devices, magnetic sensors, electric-field-controlled low-power memory devices, and spintronics technologies.[1–4] Consequently, the investigation of ME coupling has been a major focus of research in both magnetoelectric composites and single-phase multiferroic materials that exhibit both ferroelectric and magnetic order parameters.[5–8]

Among multiferroics, type-II multiferroics are particularly attractive because electric polarization originates directly from magnetic ordering, leading to an intrinsically strong coupling between electric and magnetic degrees of freedom.[9,10] However, it has been noted that in this category of materials, the ME coupling effect becomes significant only at low temperatures. More recently, low-temperature ME effects have also been inferred in low-dimensional materials such as two-dimensional van der Waals (vdW) magnets[11] and single-molecule magnets (SMMs) where magnetic bistability originates from the individual molecular unit.[12–15] These systems have therefore attracted interest for cryogenic information processing and memory applications.[16] SMMs are especially promising due to their molecular-scale size and quantum coherence properties.[17] Moreover, if the SMMs can demonstrate intrinsic ME coupling, then it would enable efficient control of molecular spin states and provide fast, localized, and energy efficient manipulation of spin qubits.[12,15]

Thus, reliable low-temperature ME characterization is essential for testing these advanced materials like vdW or SMMs in cryogenic environments. Key parameters for quantifying ME strength are given by direct measurement of the magnetic-field-induced polarization P as $\alpha_{ME} = \frac{\partial P}{\partial H}$ (direct ME coefficient), or electric-field-induced magnetization as $\alpha_{ME} = \frac{\partial M}{\partial \mathrm{E}}$ (converse ME coefficient)[18].

A straightforward approach to quantifying the direct magnetoelectric (ME) coupling coefficient is by measuring the voltage generated across the sample upon applying a magnetic field. This is most commonly implemented using a lock-in amplifier technique.[19–22] In this technique, a small, sinusoidal alternating-current (*ac*) magnetic field, typically a few Oersted (Oe), is superimposed onto a direct-current (*dc*) bias magnetic field. The resulting induced ME voltage is then measured using a lock-in amplifier locked to the excitation frequency of the *ac* magnetic field, which gives the ME coupling coefficient as:[18]

$$\alpha_{ME} = \frac{V_{lock-in}}{h_0 T} \tag{1}$$

Where, $V_{lock-in}$ is the lock-in voltage, $T$ is the thickness of the sample, and $h_0$ is the ac excitation field.

This lock-in detection scheme offers exceptionally high sensitivity; additionally, the time-varying nature of the ac field effectively mitigates charge accumulation issue which is common in polycrystalline samples. Despite these advantages, the technique is susceptible to parasitic inductive voltages that can distort the measured signals.[23] Although careful optimization of sample geometry, low-loss coaxial (BNC) cabling, and differential voltage sensing can minimize these artifacts, fully eliminating the inductive background remains a challenge.[20] Furthermore, extending these measurements to cryogenic temperatures introduces additional technical hurdles, such as spurious inductive pickup along long cryostat wiring paths and background contributions from sample conductivity, issues that have received increasing attention in recent literature.[23]

Keeping in mind the distinct technical challenges, in the present study we report a closed-cycle refrigerator (CCR)-based dynamic ME coefficient measurement setup capable of operating across 20-300 K under dc magnetic fields up to 7.5 kOe. We discuss in detail the design considerations implemented to mitigate parasitic inductive backgrounds and other cryogenic-specific artifacts. Performance of ME measurement setup is demonstrated on $CoFe_2O_4$-$BaTiO_3$ particulate composite, which have well-documented room temperature ME values [24–26] and expected temperature-dependent anomalies in ME response that correlate with the structural phase transitions of $BaTiO_3$.

## 2. Theory of the dynamic ME coupling technique

The theoretical framework for computing direct magnetoelectric (ME) coefficients originates from the Taylor series expansion of the Landau free energy density $g(E,H,T)$.[27] The magnetic-field-induced polarization in a polycrystalline can be expressed as:

$$P(H) = \frac{dg(E,H,T)}{dH} = \alpha H + \beta H^2 + \gamma H^3 \ldots \tag{2}$$

For a sample with ME coupling, in an ideal parallel-plate capacitor geometry, this polarization can be expressed in terms of electric field across the sample as:[28]

$$P \propto E = \frac{V}{T} = \alpha H + \beta H^2 + \gamma H^3 + \delta H^4 + \cdots \tag{3}$$

where $V$ is the ME voltage across the sample and $T$ is the thickness of the sample. The magnetoelectric coefficient is defined as:

$$\alpha_{ME} = \frac{dE}{dH} = \frac{1}{T}\frac{dV}{dH} = \alpha + 2\beta H + \cdots \tag{4}$$

When a small *ac* field of amplitude $h_0$ and angular frequency $\omega = 2\pi f$ is applied along with a *dc* magnetic field, the total magnetic field is given by:

$$H_{total} = H_{dc} + h_0\, sin(\omega t) \tag{5}$$

Substituting Eq. (5) into (3), the ME voltage generated across the sample:

$$\frac{V}{T} = \alpha H_{dc} + \beta H_{dc}^2 + h_0(\alpha + 2\beta H_{dc})\, sin(\omega t) + \frac{1}{2}\beta h_0^2 - \frac{1}{2}\beta h_0^2\, cos(2\omega t) \tag{6}$$

For $\frac{h_0}{H_{dc}} \ll 1$, only the first- and second-order effects are significant and the higher-order terms can be neglected. Within the first harmonic approximation, where only the $\sin(\omega t)$ term contributes, the lock-in voltage is:

$$V_{lock-in} = h_0 T(\alpha + 2\beta H_{dc}) \tag{7}$$

Comparing Eqs. (7) and (4), the magnetoelectric coefficient $\alpha_{ME}$ is obtained by:

$$\alpha_{ME} = \frac{V_{lock-in}}{h_0 T} \tag{8}$$

The Eq. (8) is valid under the first-harmonic approximation and constant phase relation. A schematic of the measurement scheme for dynamic ME coupling coefficient measurement setup is depicted in Figure 1. The setup consists of an electromagnet for generating the *dc* magnetic field, a Helmholtz coil for generating the *ac* magnetic field, an *ac* current source for driving the Helmholtz coil, and a lock-in amplifier for detecting the ME voltage across the sample.

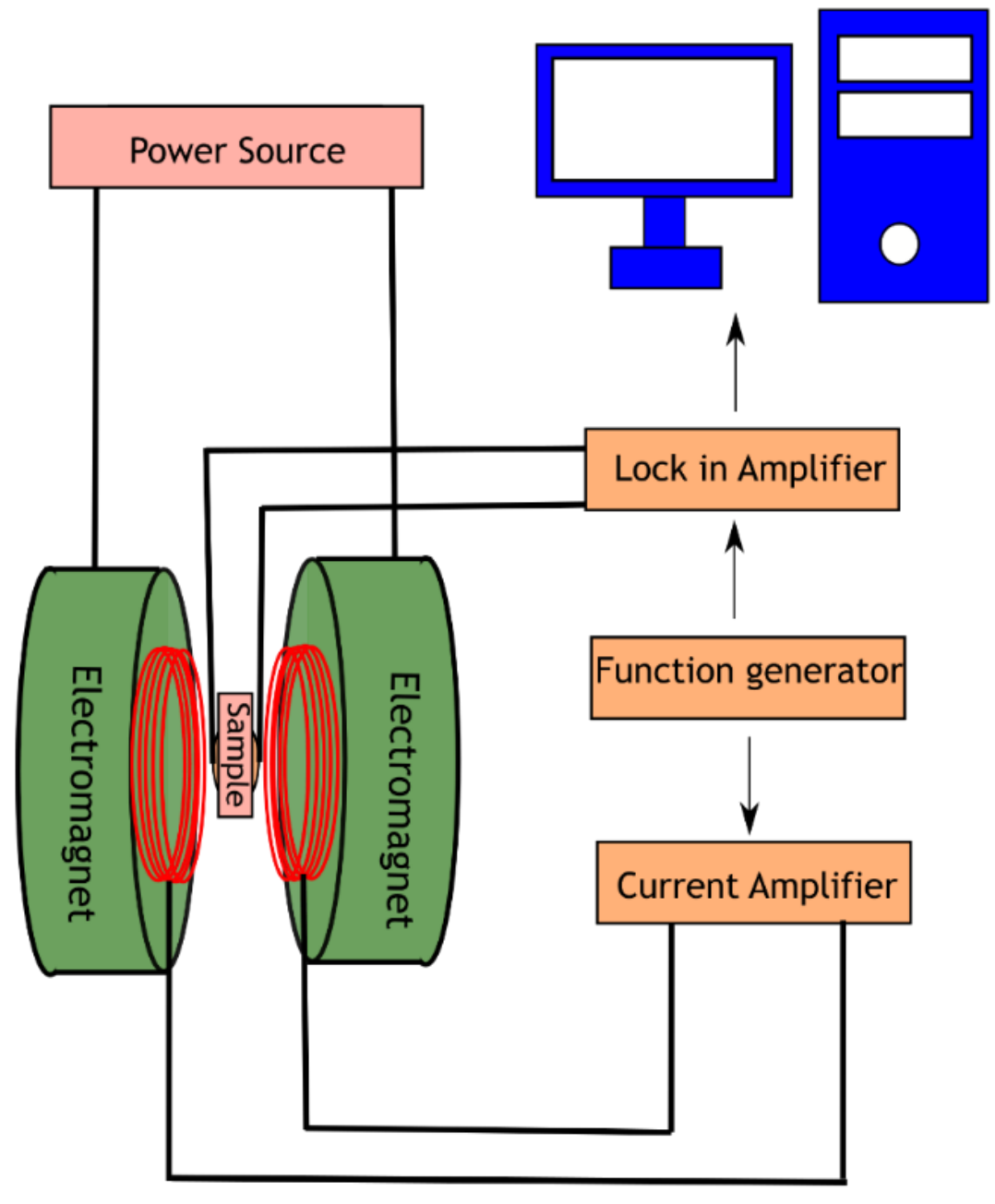

*Figure 1. A schematic of a typical lock-in technique based ME coefficient measurement setup.*

Owing to its simple implementation, dynamic method has been used amply at room temperature.[21,29–32] However, considering the distinct technical challenges (mentioned above) arising from the inductive contribution of the long wiring paths used in the cryostat, the sample's conductivity, etc., in the measured ME voltage, we propose a closed cycle refrigerator based ME coefficient measurement technique. These challenges and the design considerations that were implemented are described in detail in the next section.

## 3. Design consideration and implementation

### 3.1. ME voltage measurement by minimizing the inductive contribution

One of the common strategies for minimizing inductive voltage pickup is differential-mode voltage measurement using low-capacitance, well-shielded coaxial cables. Based on this consideration, two micro-miniature coaxial cables (1.02 mm diameter, 50 Ω impedance, silver-coated copper conductor with 0.66 mm low-dielectric PFA (Perfluoroalkoxy Alkane) insulation, and 100% aluminized Mylar shielding) specifically designed for cryogenic and UHV applications were used to connect the sample electrodes to the lock-in amplifier.

In addition to appropriate cabling, the inductive contribution to the measured signal arising from flux change, given by $V_{inductive} = \frac{d(\vec{B}.\vec{A})}{dt}$, where $\vec{B}$ is the applied ac magnetic field and $\vec{A}$ is the effective area of the measurement loop, was also considered. To minimize this contribution, the measurement cables were placed in close proximity to each other, and the configuration was aligned in a way to minimize $\vec{B}.\vec{A}$ (as shown in Figure 2(a)), thereby suppressing the inductive pickup in the measured signal.

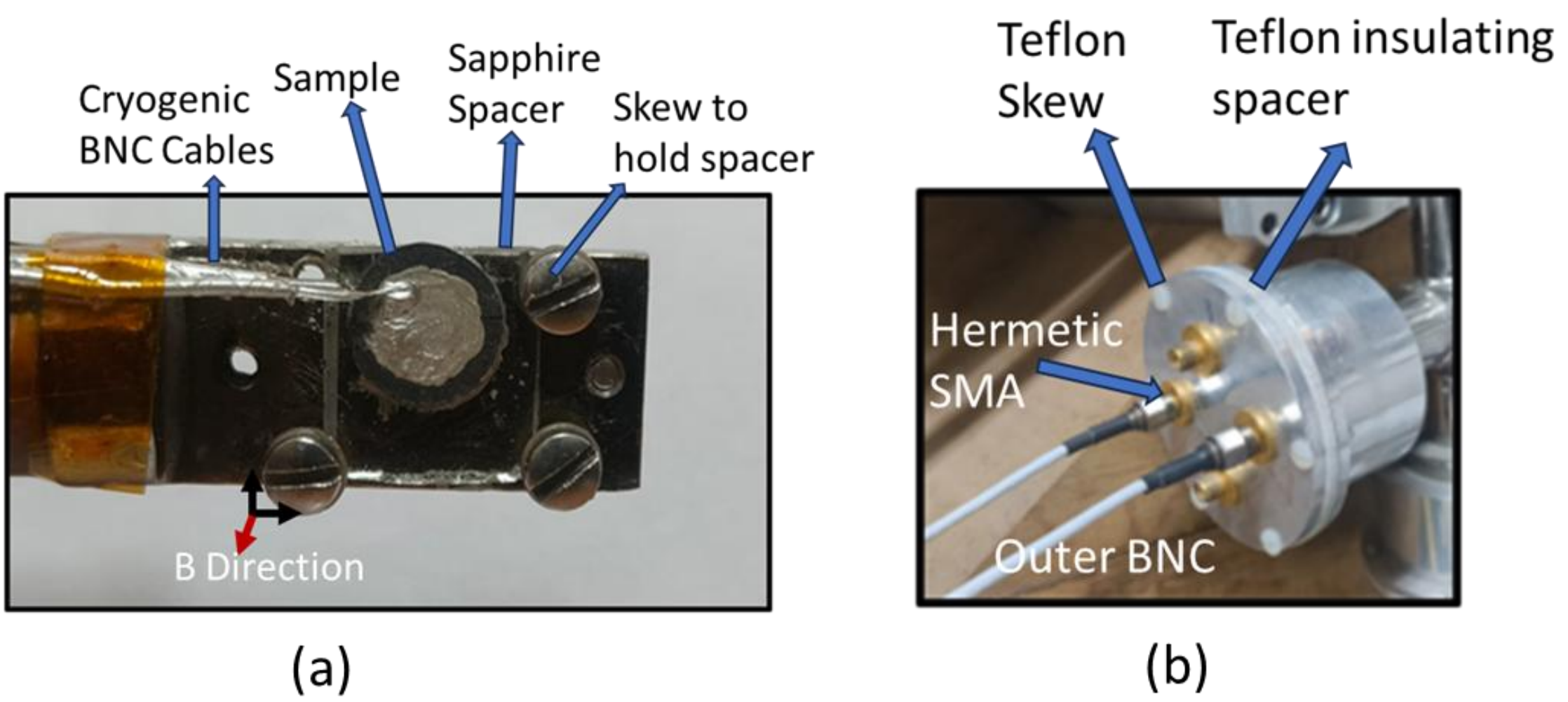

*Figure 2. (a) Sample mounting configuration at the cryo-tip sample holder. (b) Electrical feed-through attachment to the cryostat.*

### 3.2. Temperature control of the sample in the evacuated CCR chamber

For temperature-dependent ME measurement, an Advanced Research Systems (ARS) DE-202PE cryo-tip with a narrow-gap extension was utilized for sample mounting. The geometric positioning of the cryostat was such that the sample was positioned in the middle of the pole pieces of the electromagnet. The sample holder used was made of a ~3 mm thick piece of oxygen-free copper with corrosion-resistant coating for its high thermal conductivity (as shown in Figure 2(a)). A silicon diode sensor was placed on the sample holder to monitor the sample temperature. An additional temperature sensor was placed slightly away from the sample and adjacent to the heater for controlling stage temperature and minimizing the influence of magnetic fields during measurement. A Lakeshore Model-331 temperature controller was used for both monitoring and controlling the temperature. A nonmagnetic vacuum shroud was used to provide necessary vacuum insulation for cold stages and samples. The cryoprobe assembly was then inserted between the electromagnet pole pieces.

### 3.3. Electrical isolation of the measurement circuit and sample from the cryostat

For accurate measurement of ac voltage signals, measurement circuit's ground should be electrically isolated from the cryostat ground used for sample mounting. To achieve this, two isolation strategies were implemented.

(i) A sapphire spacer was placed between the sample and the cryostat cold stage to provide good electrical isolation from the grounded cryostat components while maintaining efficient thermal contact (as shown in Figure 2(a)).

(ii) The measurement instrument ground was isolated from the cryostat ground by mounting the hermetic SMA electrical feedthrough on a 2-mm thick Teflon spacer, secured with a PTFE screw, rather than directly onto the cryostat body (as shown in Figure 2(b)). This mounting scheme electrically isolated the entire measurement system from the grounds of other instruments and ensured reliable operation under vacuum conditions.

Furthermore, this configuration allows the dielectric constant measurements of the sample through a shielded two-terminal (2T) geometry employing complex impedance measurement-based methods. For this purpose, all cables and connectors used were 50 Ω impedance-matched to minimize signal loss and ensure accurate measurements.

### 3.4. Magnetic field generation

An electromagnet, Model HEM 100 (Polytronic Corporation, India) was used for generating the *dc* magnetic field in the range -7.5 kOe < H < +7.5 kOe for a pole gap of 30 mm. A programmable bipolar power supply, Model-BCS-100 (Polytronic Corporation), was used to control and drive the current in the electromagnet for *dc* field variation. A *dc* Hall probe sensor, Model DGM-204 (SES Instruments), was used for measuring the magnetic field value by placing it between the poles. For generation of the superimposing *ac* magnetic field, a Helmholtz coil (45 mm radius, 70 turns each coil) was mounted on the pole pieces. The internal oscillator of the lock-in amplifier, in combination with a power amplifier, was used to drive the ac field in the Helmholtz coil. The *ac* field value was calculated from the current measurement using a digital multimeter (DMM Fluke 87 V).

### 3.5. Measurement equipment

Lock-in amplifier model SR830 with 10 MΩ, 25 pF input impedance was used for the voltage measurement in differential mode for reducing the background induction voltage. Complex impedance measurements were conducted using an Agilent 4294A impedance analyzer, which operates on the auto-balancing bridge principle.

Combining the elements described above, the complete measurement setup is depicted in Figure 3. A LabVIEW-based program was implemented for automated control of measurement parameters and data acquisition.

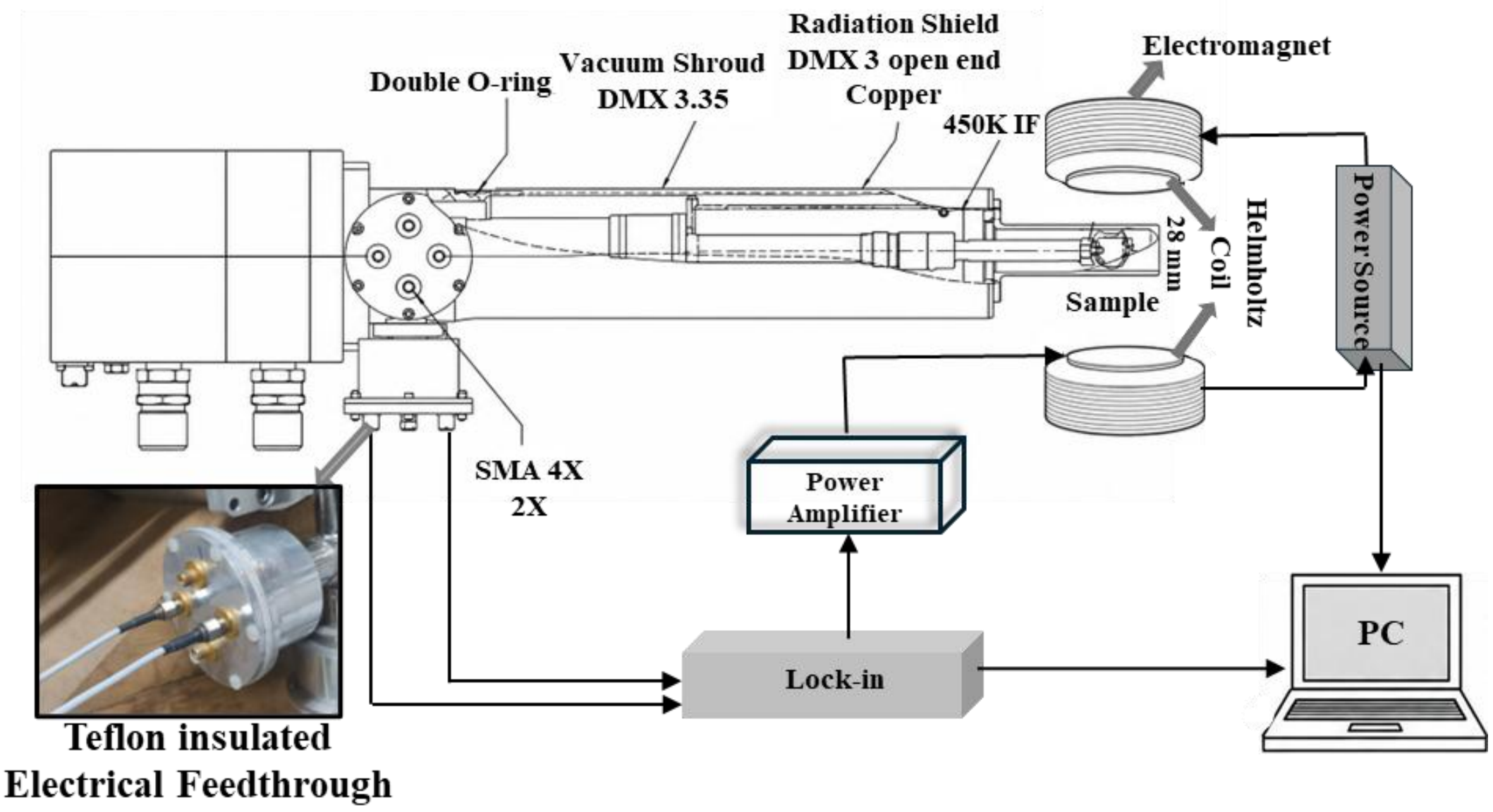

*Figure 3. Schematic of the low temperature magnetoelectric (ME) coupling measurement setup.*

## 4. Instrument validation and performance characterization

To validate the measurement setup, a model $CoFe_2O_4$-$BaTiO_3$ (CFO-BTO) particulate composite was characterized using the custom-built setup described above. CFO-BTO is a well-studied strain-mediated magnetoelectric (ME) system with reported reference values for ME coupling.[25,26] In this composite, an applied magnetic field induces magnetostrictive strain in the CFO phase, which is transferred across the CFO-BTO grain boundaries to the piezoelectric BTO phase, generating a measurable ME voltage.

CFO is a ferrimagnetic material with a Curie temperature of ~793 K, far above the present measurement range.[33] Thus, the magnetic ordering of the CFO phase remains stable throughout the measurement range of 20-300 K, with no associated structural transition expected in this temperature interval. Its room-temperature and low-temperature magnetostrictive properties are also available in the literature over the range of 10-300 K.[34,35]

The BTO component is also a well-known ferroelectric material exhibiting structural phase transitions that are well-documented.[36–38] On cooling from room temperature (~300 K), BTO undergoes a tetragonal-to-orthorhombic transition near 280 K, followed by an orthorhombic-to-rhombohedral transition near 200 K. These transition temperatures are also retraced, with a small thermal hysteresis, upon warming from the lowest measurement temperature back to room temperature (see curve (a) in Figure 11).[38]

Thus, we chose to study this composite as a comprehensive benchmark for validating the custom-built low-temperature M-E coupling measurement setup developed in our lab. For this purpose, we took the following steps:

(i) Since CFO is the sole electrically conducting phase, identical measurements on a pure CFO pellet serve as a null/reference to identify inductive voltage contribution, and for baseline check on spurious artifacts if any, present across the full temperature range.
(ii) The sensitivity and data resolving capabilities of the instrument are demonstrated by setup's ability to resolve subtle ME coupling anomalies corresponding to the dielectric phase transitions of BTO on the measured temperature scale.
(iii) Finally, absolute measurement accuracy can be established by benchmarking the room-temperature ME coupling coefficient against literature reference values.

### 4.1. Parasitic background from inductive voltage

To evaluate the contributions from parasitic inductive voltage in our measurement setup, we measured a pure CFO pellet, which represents the conductive constituent of the CFO-BTO composite. Since no genuine ME voltage is expected from pure CFO, any detected signal can be attributed primarily to the inductive contribution. Figure 4(a) presents the lock-in voltage measured at 1 Oe with ac excitation frequency variation. The voltage increases with increasing frequency, which is characteristic of an inductive voltage. Although the background signal is relatively small, about 3.5 $\mu$V at 991 Hz, it becomes very significant when the intrinsic ME coupling is weak.

The magnitude of the inductive voltage depends on the parameters of the applied *ac* magnetic field, whereas the *dc* magnetic field is not expected to contribute appreciably to this background response. This behavior is confirmed by the *dc*-field dependence shown in Figure 4(b), where the lock-in voltage remains nearly constant over the applied *dc* magnetic field range under fixed *ac* excitation conditions.

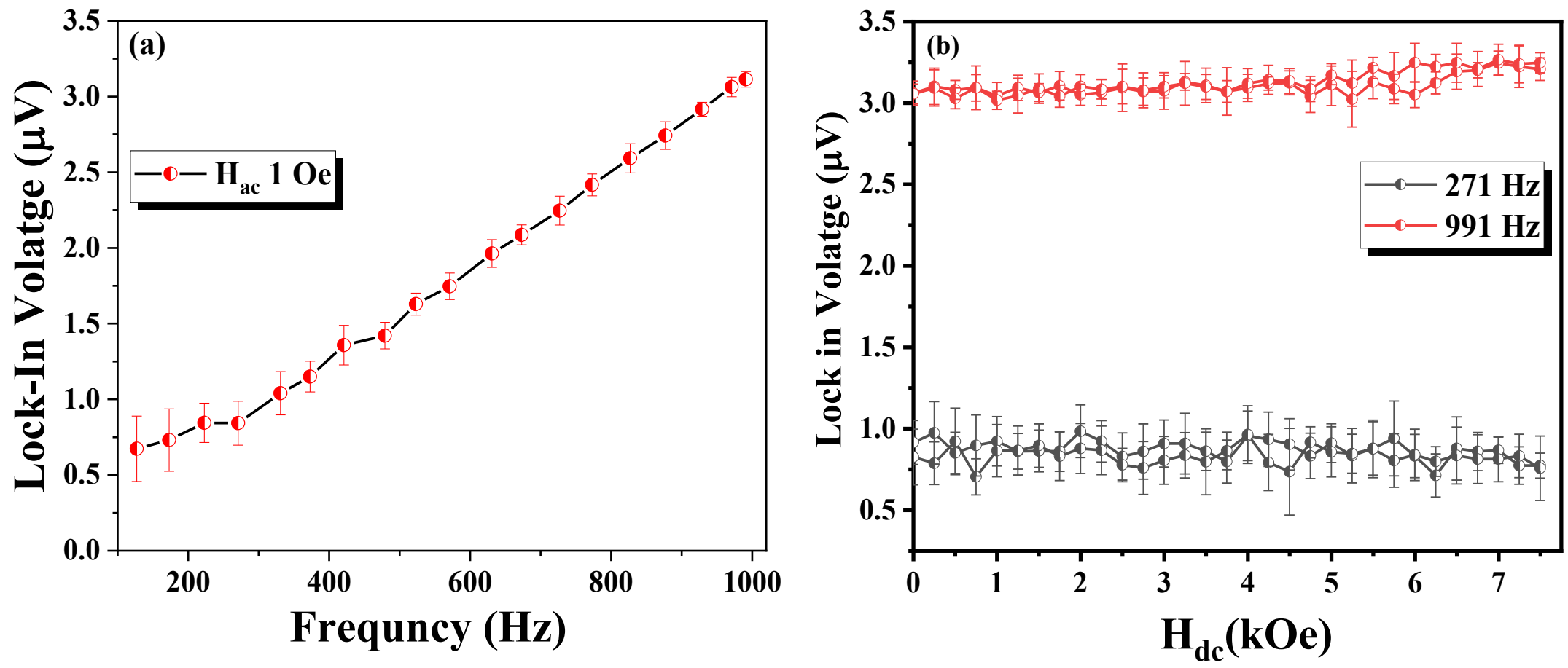

*Figure 4. Frequency and dc-field dependence of the parasitic background voltage measured on a pure CFO pellet. (a) Lock-in voltage as a function of ac excitation frequency (100-1000 Hz) at $H_{ac}$ = 1 Oe, showing the near-linear increase characteristic of an inductive background. (b) Lock-in voltage as a function of dc bias field (0-7.5 kOe) at fixed frequencies (271 Hz and 991 Hz), showing no significant dc-field dependence.*

### 4.2. Dielectric measurement calibration

Accurate characterization of the dielectric properties of the ME composite is essential for correlating the ME response with dielectric anomalies at the BTO structural transitions. The dielectric properties were determined from capacitance measurements using complex impedance spectroscopy of the sintered composite pellets. The measurement system was calibrated as follows:

#### *4.2.1. Impedance analyzer accuracy*

The accuracy of the Agilent 4294A impedance analyzer was verified using a standard 100 Ω resistor (Agilent 04294-61001) over 40 kHz to 1 MHz. The measured resistance values consistently remained within 0.1 % of the nominal value, confirming the analyzer's accuracy.

### *4.2.2. Capacitance measurement accuracy*

The capacitance measurement setup was validated by measuring a low loss commercial ~91 pF capacitor, with measurements compared against the standard Agilent 16047A fixture. As shown in Figure 5 (Top), both methods agreed within 0.5% below 500 kHz. The deviation increases gradually above 500 kHz, attributed to additional cable length and connector parasitics (see Figure 5 (Bottom)). The measured dissipation factor remained below $2\times10^{-3}$, confirming the suitability of the setup for accurate dielectric characterization over the investigated frequency range (see Figure 6).

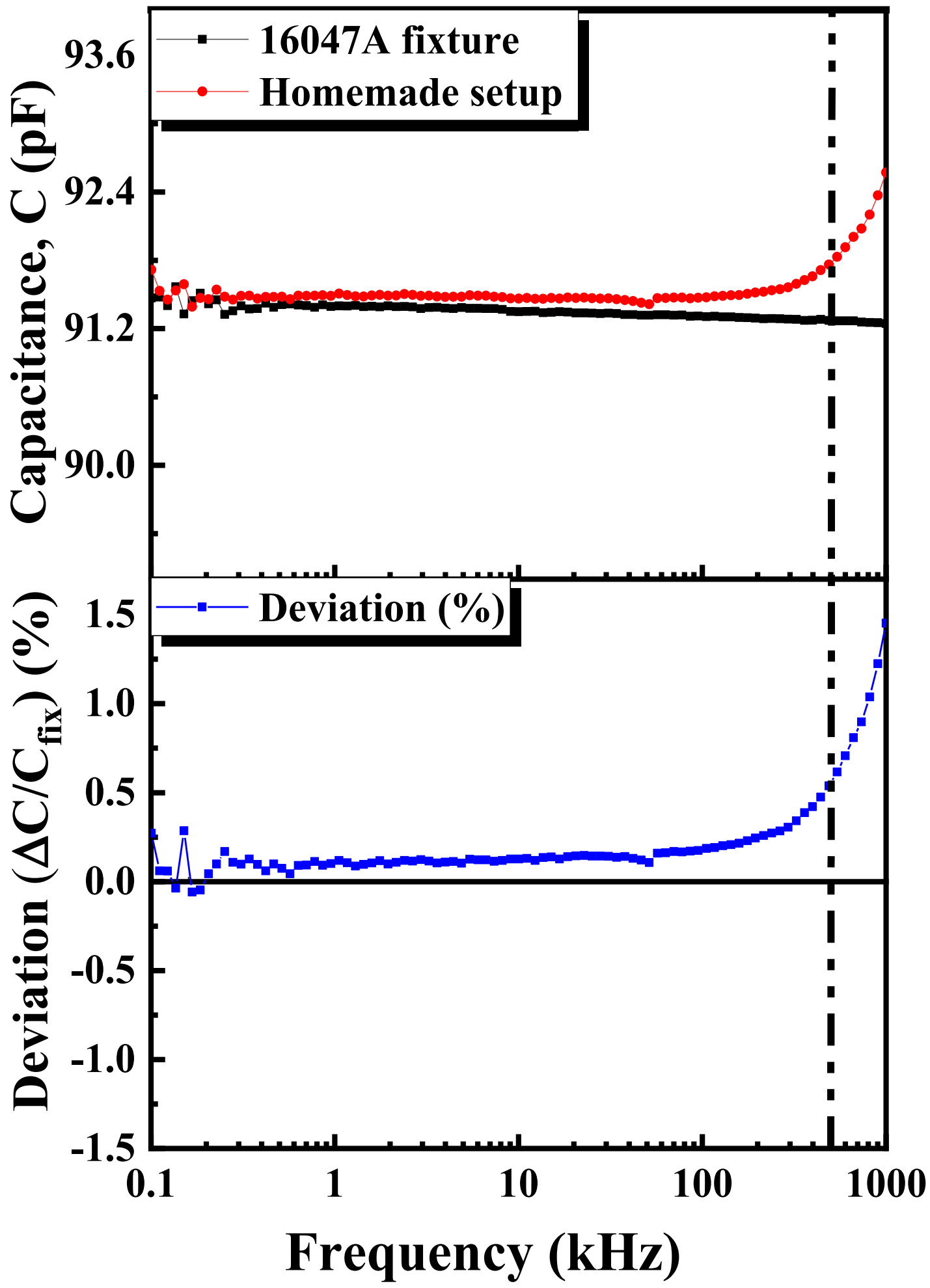

*Figure 5. (Top) Measured capacitance (C) as a function of frequency from 100 Hz to 1 MHz for low loss commercial capacitor comparing the standard fixture (Agilent 16047A) and the home-built setup. (Bottom) Relative deviation percentage between the two measurement setups. The vertical line marks 500 kHz below which the discrepancy remains under ~0.5 %.*

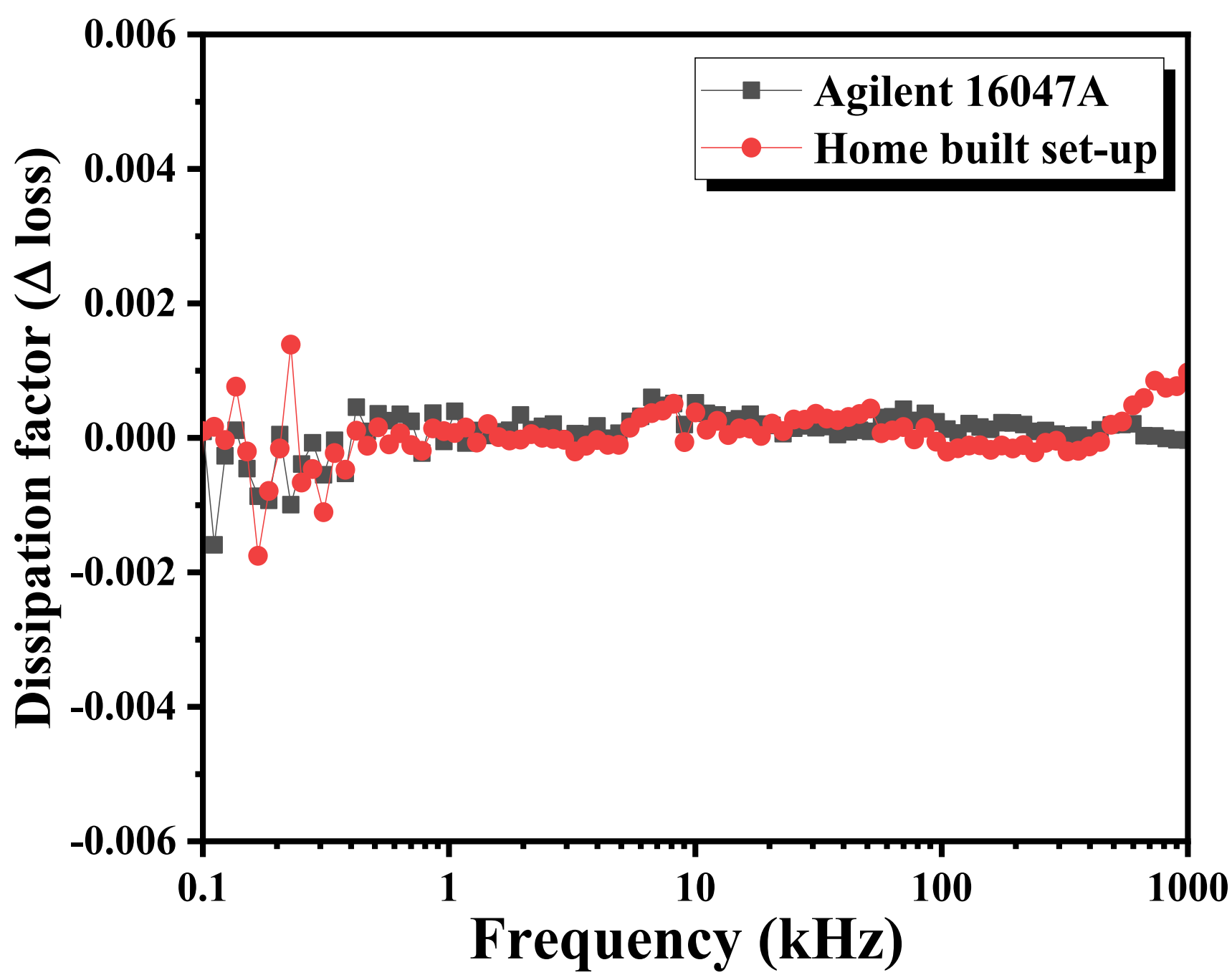

*Figure 6. Dissipation factor vs. frequency (100 Hz-1 MHz) for the low-loss commercial capacitor, measured using the home-built setup and the Agilent 16047A fixture.*

#### *4.2.3. Temperature-dependent dielectric background*

To evaluate the background behavior under temperature-dependent measurement conditions, an empty (air) capacitor was constructed by mounting two copper electrodes (6 mm diameter, ~0.7 mm thickness) parallel to each other, fitted in a plastic straw (AGC2, Quantum Design). This capacitor was mounted onto the sample holder using silver paint and measured as a function of temperature. As shown in Figure 7, the capacitance exhibits a monotonic variation with temperature, attributed to thermal contraction of the capacitor geometry on cooling. Furthermore, the measured loss remains constant at ~$20\times10^{-3}$ across the entire temperature range, without any anomaly.

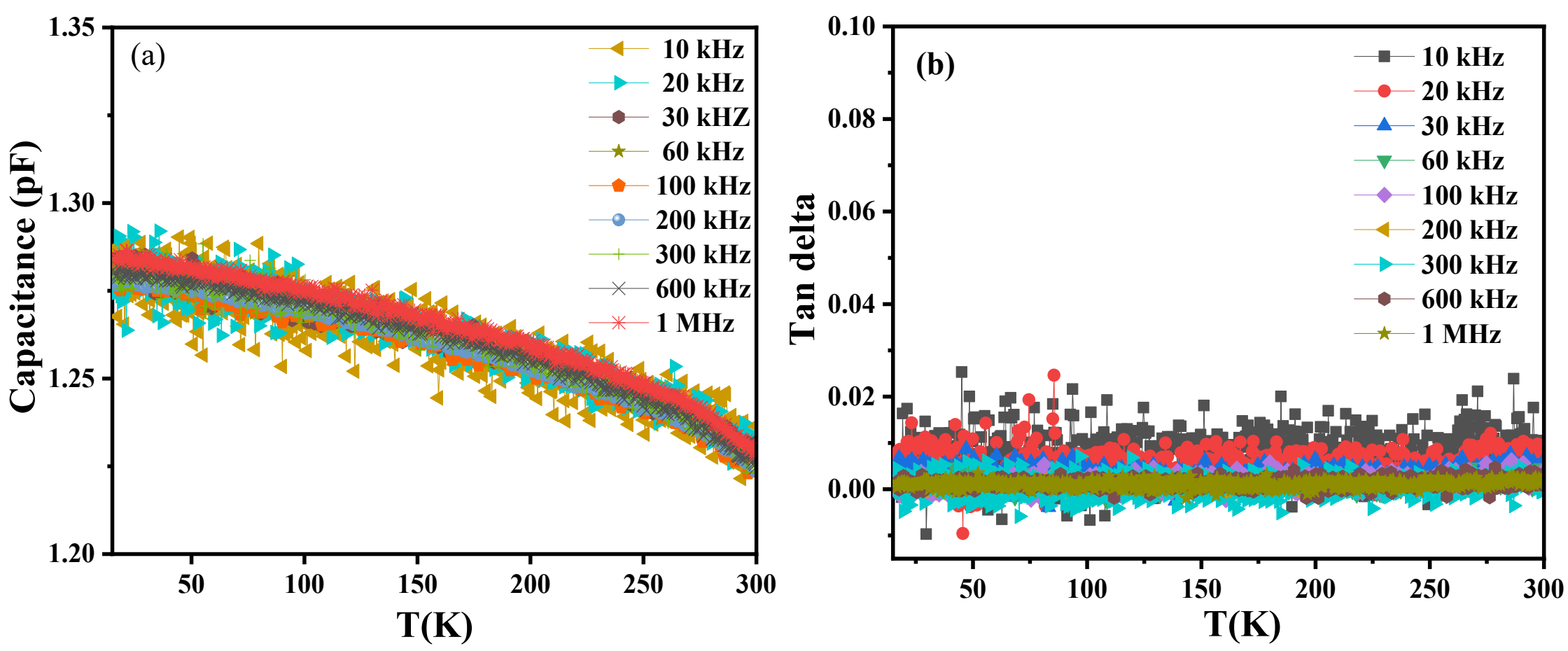

*Figure 7. Temperature dependence of (a) capacitance and (b) loss tangent (tan δ) of an empty capacitor, measured at multiple frequencies.*

### 4.3. Room-Temperature Magnetoelectric Response

To verify the proper functioning of the ME measurement system, the room temperature ME response of the CFO-BTO composite was first recorded. Measurements were performed in longitudinal geometry $H_{ac} \| H_{dc}$ using an *ac* field $H_{ac}$ = 2 Oe at 271 Hz. The room-temperature ME voltage of the CFO-BTO composite is presented in Figure 8, displaying the characteristic butterfly-shaped loop.

Conventionally, in the literature $V_{ME}$ is reported as $V_{total} = V_z = \sqrt{V_x^2 + V_y^2}$ and the phase angle is given by $\theta = \arctan(V_y/V_x)$, where $V_x$ and $V_y$ are the in-phase and out-of-phase voltage components, respectively.[19,20] However, as pointed out by Amiri et al., the Y component of the $V_{ME}$ can be affected by the inductive voltages coming from the limited conductivity if any, of the sample.[23] However, as discussed in Section 4.1, the inductive contribution remains small at low frequency and this low field amplitude, and is largely independent of the *dc* magnetic field.

Furthermore, it should be noted that the *Y* component of the $V_{ME}$ can also appear from the inductive-resistive (LR) characteristics of the Helmholtz coil that give rise to a phase lag between the applied *ac* magnetic field and the reference signal of the lock-in amplifier that varies with frequency. Also, the frequency-dependent phase shift introduced by the Helmholtz coil and the power amplifier makes an absolute phase difficult to establish. Therefore, the in-phase, out-of-phase, and total ME signal are reported together (Figure 8). Although the out-of-phase component in our measurement is small, it follows a *dc* field dependence similar to that of the in-phase component.

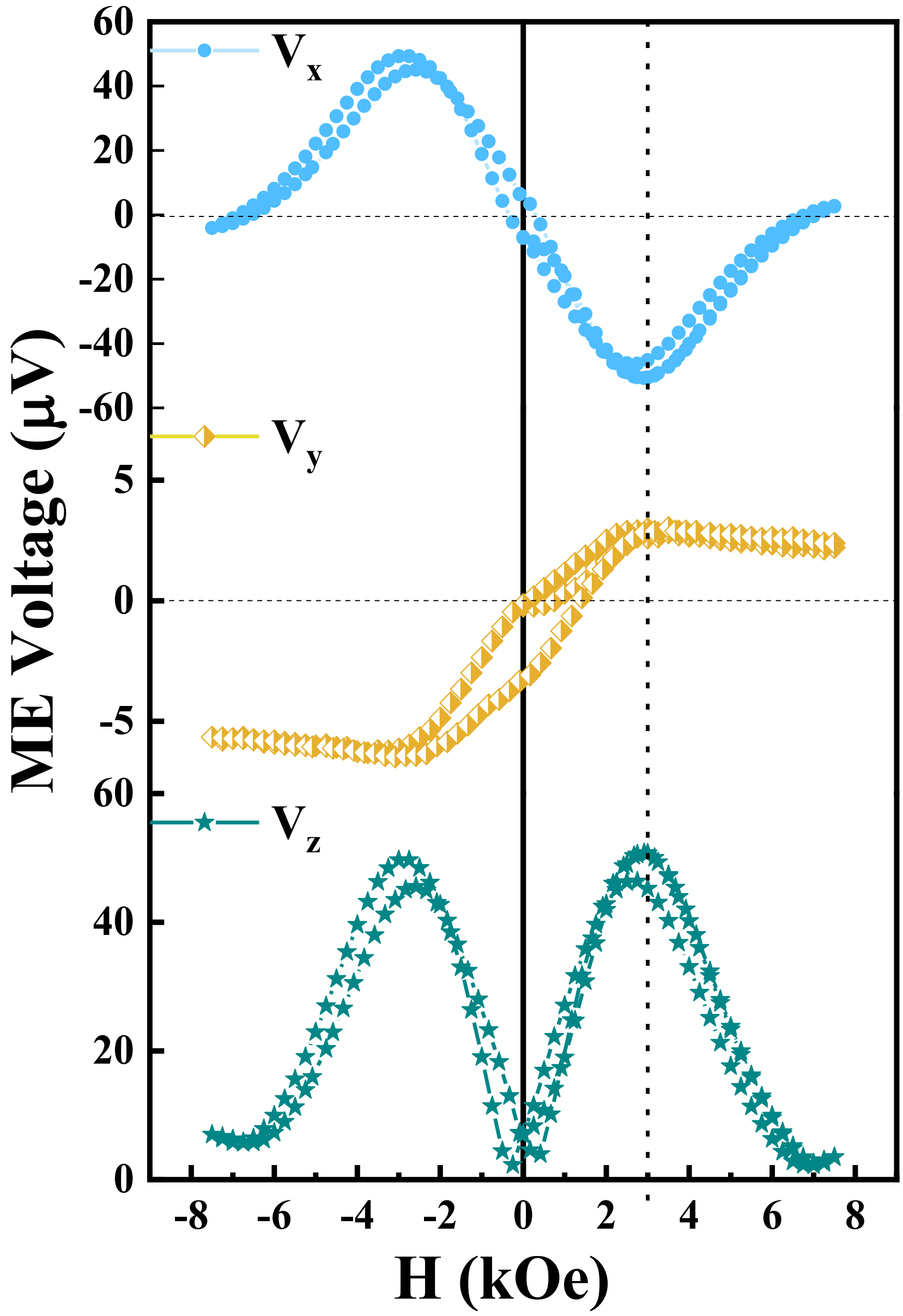

*Figure 8. In-phase ($V_x$), out-of-phase ($V_y$), and total ($V_z$) ME voltage of the CFO-BTO composite as a function of dc magnetic field at 300 K, measured at Hac = 2 Oe, 271 Hz.*

Furthermore, as shown in Figure 9, the measured signal values are significantly higher than the background voltage obtained for pure CFO over the investigated magnetic field range. This confirms that the instrument can reliably resolve the genuine ME signal against a low background voltage.

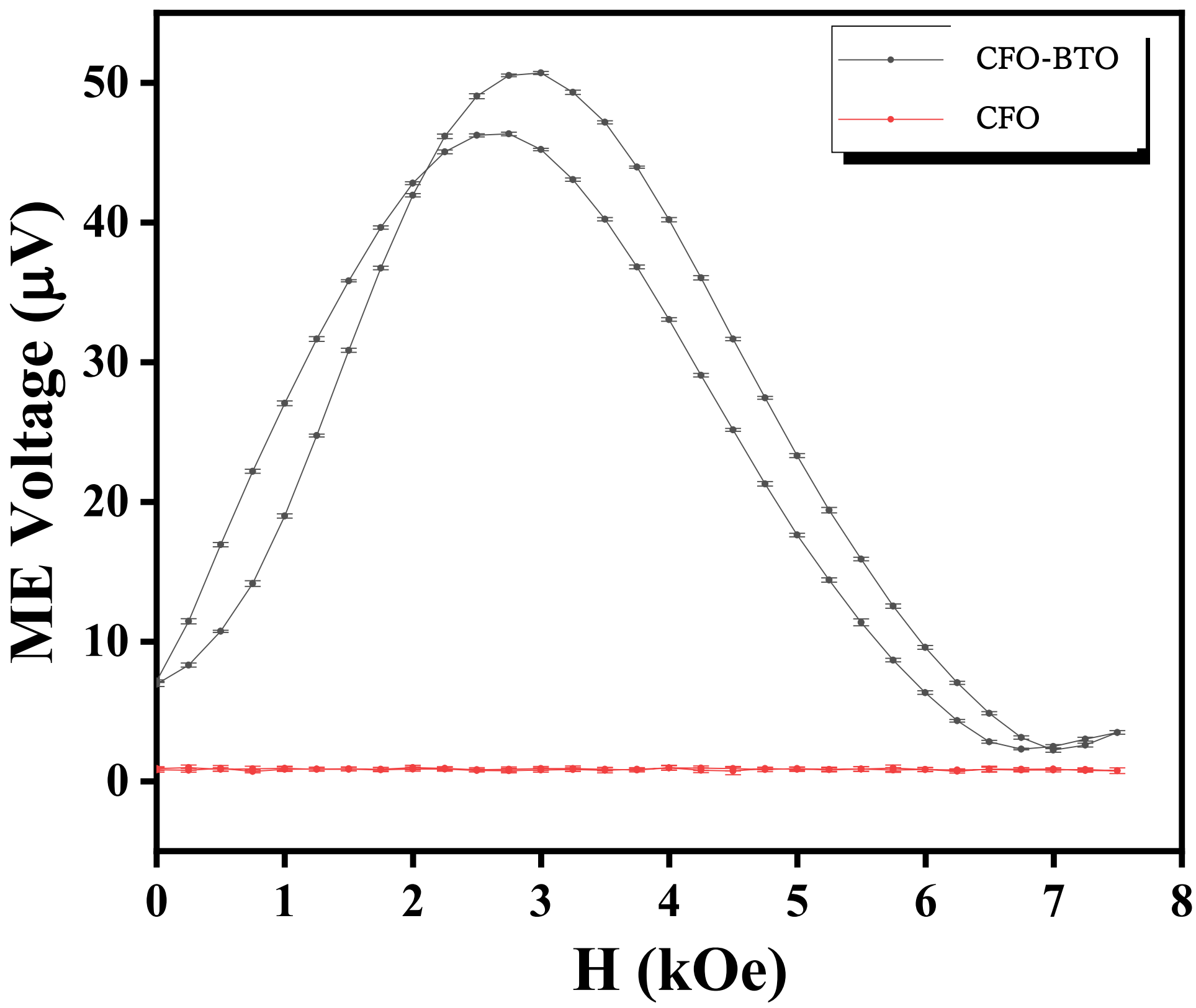

*Figure 9. Longitudinal ME voltage (total $V_z$) of $CoFe_2O_4$-$BaTiO_3$ powder mix composite, ac magnetic field: 271 Hz, 2 Oe.*

The calculated values of maximum ME coefficient of 0.23 mV/cm·Oe was observed at a *dc* bias field of 3000 Oe, is in good agreement with values reported for CFO-BTO powder-mixed composites, confirming the reproducibility of the ME coefficient values obtained with our setup.[26]

### 4.4. Temperature-dependent ME coupling

The temperature-dependent magnetoelectric (ME) coupling response of the CFO-BTO composite was recorded over the temperature range of 20-300 K. The isothermal ME voltage responses measured at different temperatures are shown in Figure 10. Across all temperatures displaying a significant magnetoelectric response, the response exhibits a maximum at approximately 3 kOe. At 50 K, the ME coupling remains low and no characteristic maximum behavior has been observed for the $H_{dc}$ field sweep. With increasing temperature, the maximum behavior appears above 200 K and increases monotonically with temperature for in-phase and the total voltage component. However, out-of-phase component has its highest value at near 280 K and then the value decreases at 300 K.

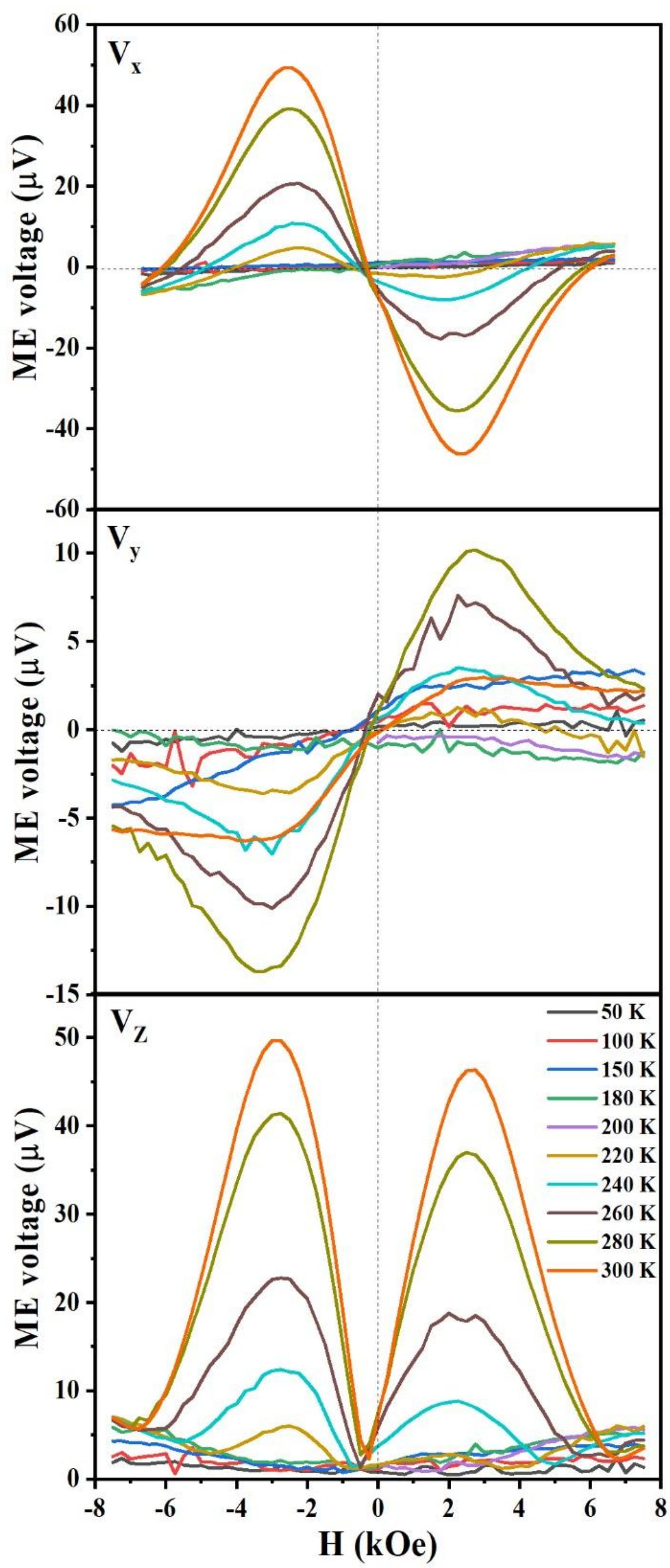

*Figure 10. Temperature-dependent ME voltage [(top) in-phase $V_x$, (middle) out-of-phase $V_y$, (bottom) total $V_z$] of the CFO-BTO composite as a function of dc magnetic field, measured from 50-300 K.*

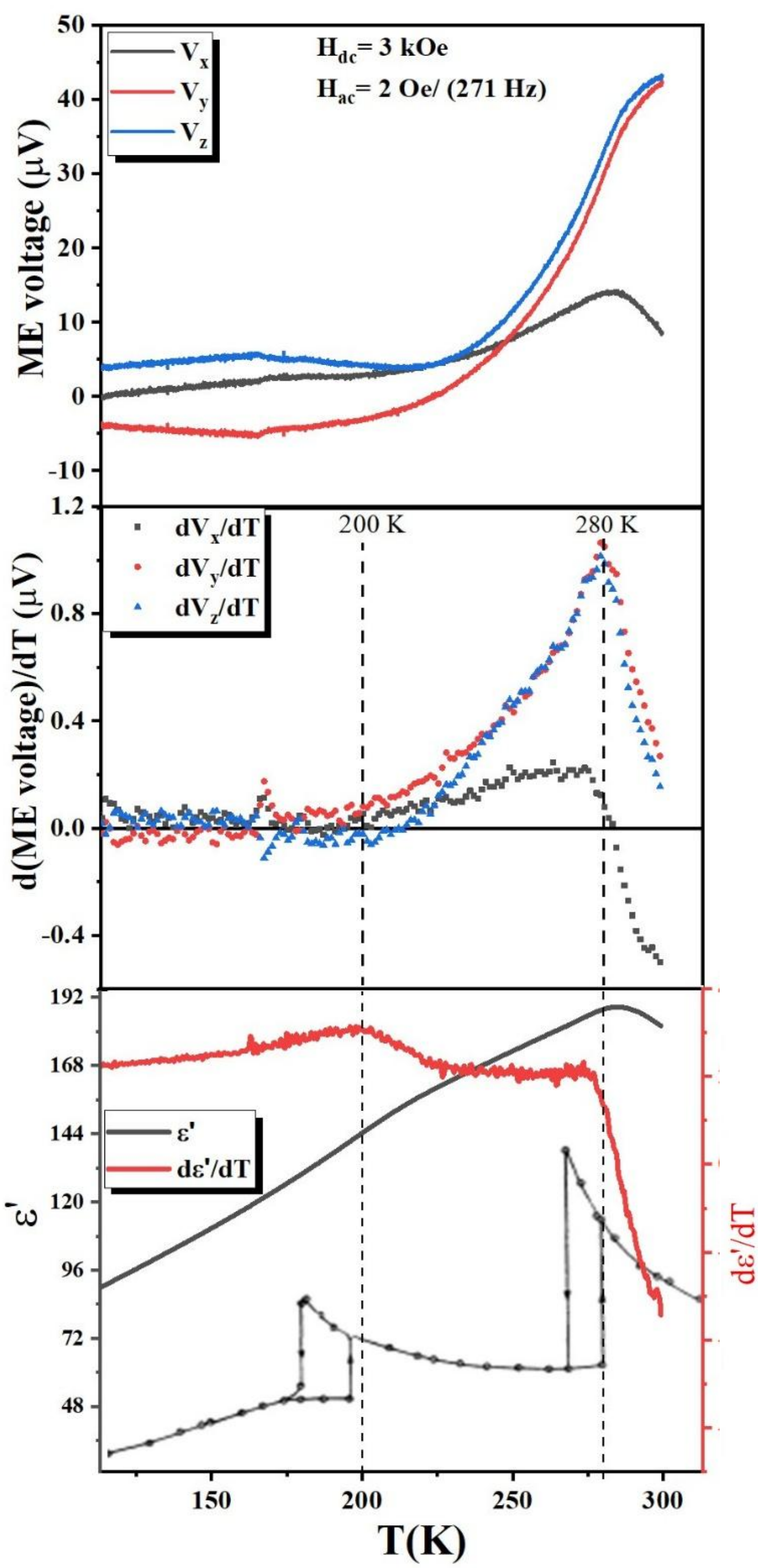

*Figure 11. (Top) In-phase $V_x$, out-of-phase $V_y$ component and total ME voltages $V_z$. (Mid) Derivative of the In-phase $V_x$, out-of-phase $V_y$ component and total ME voltages $V_z$ (Bottom) Dielectric constant of CFO-BTO pellets at 100 kHz and its derivative compared with the literature.*[38]

Across all temperatures displaying a significant magnetoelectric response, the response exhibits a maximum at approximately 3 kOe. At 50 K, the ME coupling remains low and no characteristic maximum behavior has been observed for the $H_{dc}$ field sweep. With increasing temperature, the maximum behavior appears above 200 K and increases monotonically with temperature for in-phase and the total voltage component. However, out-of-phase component has its highest value at near 280 K and then the value decreases at 300 K. To further characterize this behavior, we performed temperature-sweep measurements at a constant field of 3 kOe; all three components are shown in Figure 11. The derivatives of the in-phase, out-of-phase, and total ME voltage components each reveal anomalies near 200 K and 280 K. The corresponding background signal of CFO remained constant and negligible relative to the CFO-BTO composite signal.

These temperatures are in close agreement with the known dielectric transition temperatures of BaTiO3, corresponding to the rhombohedral-orthorhombic and orthorhombic-tetragonal phase transitions, respectively.[38] To further validate this correspondence, the dielectric properties of the same sample were measured without altering the cryostat's internal measurement configuration, and the resulting dielectric data show corresponding anomalies at the same temperatures.

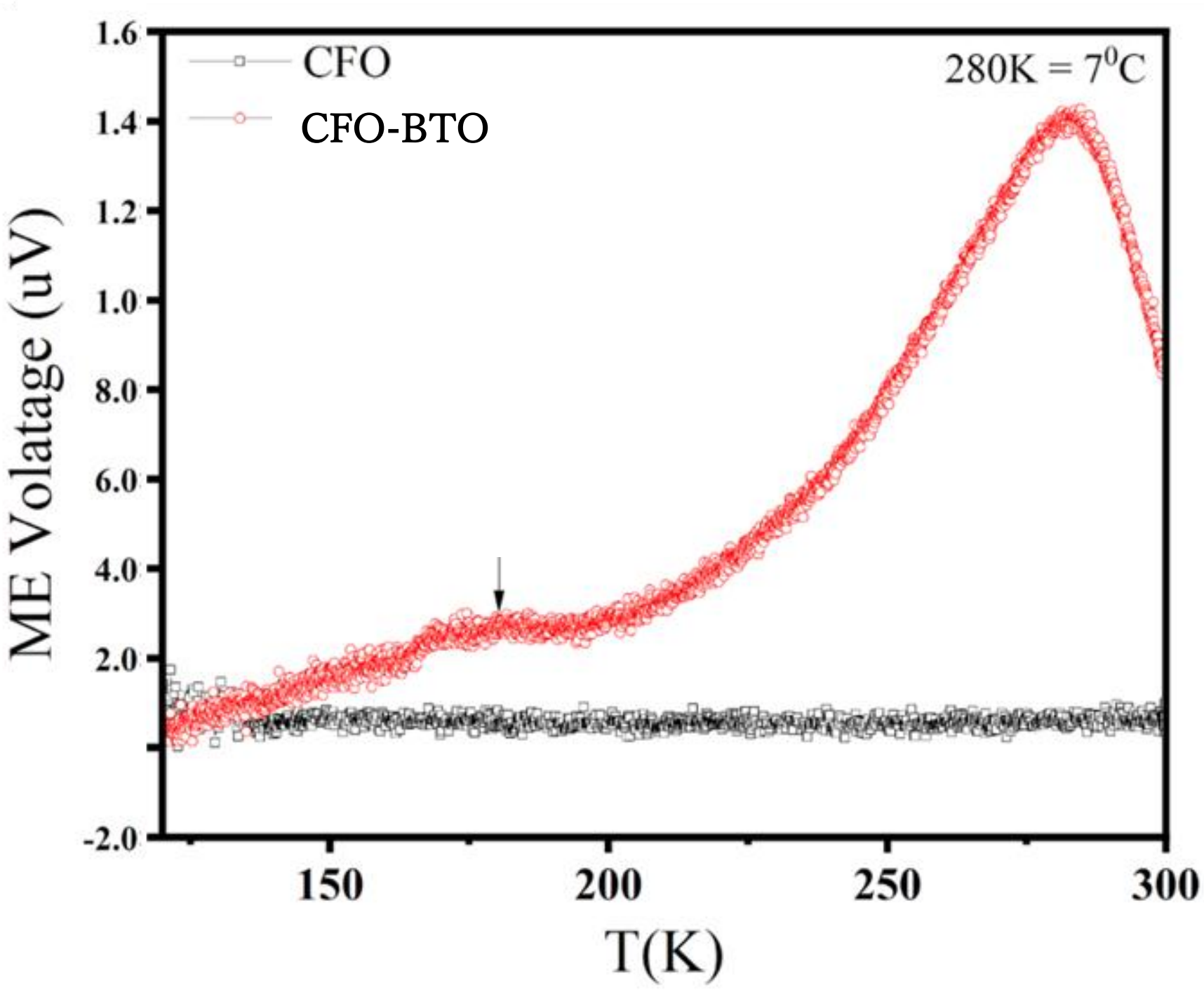

*Figure 12. Temperature dependence of the in-phase ME voltage component $V_x$ of CFO-BTO composite, compared with the CFO reference background, showing that the background remains nearly unchanged even relative to the lower-valued ME voltage component.*

Additionally, as shown in Figure 12, the measured CFO background signal remains constant across the full temperature range (20-300 K) when compared to the CFO-BTO composite data. This confirms that the instrument background does not introduce spurious features into the temperature-dependent ME measurements.

Collectively, these results validate the instrument's ability to reliably measure magnetoelectric (ME) coupling responses in the 100 Hz-1 kHz range and dielectric properties in the 100 Hz-500 kHz range across the temperature range 20-300 K and under magnetic fields up to 7.5 kOe. It has been noted that the ME coupling phase relationship also changes across the transition. Hence, instead of considering the in-phase component of the ME coupling, considering all the values $V_x$, $V_y$ and $V_z$ is more accurate for investigating the phase transition in materials.

## 5. Conclusion

In summary, we developed a closed-cycle refrigerator-based low-temperature setup for measuring magnetoelectric coupling coefficient using dynamic lock-in detection over 20-300 K, with *dc* magnetic fields up to 7.5 kOe and *ac* excitation field up to 5 Oe at 100 Hz and 1 kHz. Measurements on $CoFe_2O_4$ confirm that parasitic inductive background remains small and largely independent of *dc* field, enabling reliable background correction. The capabilities of the setup were demonstrated using a $CoFe_2O_4$-$BaTiO_3$ composite, which exhibits expected butterfly-shaped ME response and temperature-dependent anomalies that correlate with features in its dielectric permittivity. Carefully engineered cryogenic wiring and impedance-matched cabling, enable measurement of ME and dielectric properties on the same sample without reconfiguration. This versatile setup is ideal for probing weak ME coupling, detecting subtle phase transitions, and temperature-dependent coupling in multiferroic composites and quantum materials.

## Acknowledgement

The authors wish to thank the Aerospace Research & Development Board for financial support through grant ARDB/01/2031992/M/I/1994 (RP04147G). We acknowledge the CRDAM, IMR, Tohoku University for the support by GIMRT Proposal 202503-CRKKE-0510.